\documentclass[aps,twocolumn]{revtex4}
\usepackage{epsfig}
\usepackage{psfrag}
\newcommand{\be}{\begin{equation}}
\newcommand{\ee}{\end{equation}}
\newcommand{\ba}{\begin{eqnarray}}
\newcommand{\ea}{\end{eqnarray}}

\begin{document}

\title{Local persistence in directed percolation}

\author{Peter Grassberger}
\affiliation{John-von-Neumann Institute for Computing, Forschungszentrum J\"ulich,
D-52425 J\"ulich, Germany\\ and \\ Department of Physics and Astrophysics, University 
of Calgary, Alberta, Canada T2N 1N4}

\date{\today}

\begin{abstract}
We reconsider the problem of local persistence in directed site percolation. We 
present improved estimates of the persistence exponent in all dimensions from $1+1$ to 
$7+1$, obtained by new algorithms and by improved implementations of existing ones.
We verify the strong corrections to scaling for $2+1$ and $3+1$ dimensions found 
in previous analyses, but we show that scaling is much better satisfied for very 
large and very small dimensions. For $d > 4$ ($d$ is the spatial dimension), the 
persistence exponent depends non-trivially 
on $d$, in qualitative agreement with the non-universal values calculated 
recently by Fuchs {\it et al.} (J. Stat. Mech.: Theor. Exp. P04015 (2008)).
These results are mainly based on efficient simulations of clusters evolving under 
the time reversed dynamics with a permanently active site and a particular 
survival condition discussed in Fuchs {\it et al.}. These simulations suggest
also a new critical exponent $\zeta$ which describes the growth of these clusters
conditioned on survival, and which turns out to be the same as the exponent, $\eta+\delta$   
in standard notation, of surviving clusters under the standard DP evolution.
\end{abstract}

\maketitle

\section{Introduction}

In general, persistence \cite{Bray,Majumdar,Ray} in time-dependent critical phenomena is
the probability that some observable does not cross its long-time expectation until some 
finite time $t$. In particular, we will deal with the order parameter, which in directed 
percolation is the density and satisfies $\langle \rho\rangle \to 0$ at the critical 
point. Local persistence $P({\bf r},t)$ is the probability that the local order  
parameter at position ${\bf r}$ does not cross this value up to time $t$. In 
directed percolation it is 
thus equal to zero when the site ${\bf r}$ was active in the initial configuration, 
while it is positive and equal to the chance that it is not yet activated at $t$
when it was inactive originally. In the following we shall only consider homogeneous 
systems in which case $P({\bf r},t)$ does not depend on ${\bf r}$ and will be written 
$P(t)$.

In general, local persistence decays according to a power law
\be
   P(t) \sim t^{-\theta}
\ee
where $\theta$ is a new universal critical exponent, independent of the standard 
exponents. In particular, it is in general not related to the dynamical critical 
exponent $z$ \cite{Bray,Majumdar,Ray,Fuchs}.

Previous studies of persistence (in the following we will only deal with {\it local}
persistence) in the directed percolation (DP) universality class \cite{Fuchs,Hinrichsen,Albano}
have given somewhat contradictory results. In particular, {\it superuniversality} was 
suggested in \cite{Albano}, i.e. the possibility that $\theta$ is independent of 
dimension. This was later refuted in \cite{Fuchs}. For $d=1$, a connection with 
spreading of DP clusters in the presence of a wall wall was pointed out in \cite{Hinrichsen}, 
although no direct relation between $\theta$ and critical boundary exponents for DP 
could be established.

In the following we shall again study this problem by means of numerical simulations. 
In spite of doubts concerning the universality of $\theta$ \cite{Fuchs}, we restrict 
ourselves to site percolation on simple hyperbolic lattices. We 
use a new sampling strategy which is more time efficient than the standard strategy in 
low dimensions, and more space efficient in high dimension. In addition, we also made
(for intermediate dimensions) simulations with a highly efficient implementation of the 
standard strategy. Finally, since it was crucial to simulate exactly at the critical 
point, we made new estimates of $p_c$ (the percolation threshold) in dimensions 1+2 to 
1+7. One of our findings is that $\theta$ not only is not superuniversal, but it depends
also non-trivially on dimension above the critical dimension for DP (which is $d_c=4$; 
in the following, we will always denote by $d$ the spatial dimension). The latter had 
been predicted in \cite{Fuchs}, where it was also suggested that $\theta$ might be not 
universal at all for $d>d_c$.

Our new sampling strategy suggests immediately a new critical exponent, called $\zeta$
in the following. This exponent describes the growth of clusters under a modified 
evolution, where time is reversed, a permantent source of activity is added at the site
at which persistence is measured, and a rule is added that clusters die when they hit 
the baseline \cite{Hinrichsen,Fuchs}. The growth of the mass of such clusters, conditioned 
on their survival, can be described by an exponent $\zeta$. As pointed out by H. Hinrichsen 
after this paper was written \cite{HH}, a heuristic argument relates $\zeta$ to a combination of 
standard DP exponents which can be re-written, using hyperscaling and the standard 
notation for DP exponents \cite{Haye}, as 
\be
   \zeta = \eta+\delta.
\ee
The r.h.s. is the exponent which describes the mass growth of surviving critical DP
clusters under normal (forward time, single site seed) dynamics.

\section{Algorithms and the exponent $\zeta$}

In all simulations we used helical boundary conditions, i.e. a site is indexed by a 
single integer $i\in [0,\ldots N-1]$, where $N$ is the size of the lattice. In all 
cases, $N$ is chosen as a power of 2. Neighbors
of $i$ are indexed by $i\pm 1, i\pm L_1, \ldots i\pm L_{d-1}$, all modulo $N$. The 
integers $L_k$ are of order $N^{1/d}$, but not necessarily equal to it (i.e. the 
lattice does not have to be an exact hypercube).

The standard way to simulate persistence uses four data structures: 
\begin{itemize} 
\item A bit (or, for implementation simplicity, byte) array of size $N$, representing 
the activity pattern ($s_i = 1$ if site $i$ is active, $s_i = 0$ otherwise), which 
is replaced at each time step by a new empty array. It is used for checking whether 
a site to be activated is already active or not.
\item A list of indices of all active sites during the previous time step. This list 
is used to activate neighbors in the next time step. 
\item A similar list containing the active sites during the present time step, which 
originally is also empty and which will replace, after the time step is finished, 
the previous list.
\item A bit array of size $N$ representing the history of all sites: $t_i = 1$, if 
site $i$ had been active during any previous time step, and $t_i = 0$ otherwise. 
This array is updated continuously, but erased only at the start of a new run.
\end{itemize}

The simulation proceeds then in the usual way, starting with the initial list of 
active sites (which can be a single site or a finite fraction of all sites) and 
simulating time step after time step, until either a maximal time is reached, 
all activity has died out, or until all sites have been activated at least once. 
For runs which start with a single active site, only a tiny fraction of the lattice 
will actually be activated when $d$ is large. In that case a direct implementation of the bit arrays 
is wasteful of memory and is replaced by hashing. Details are discussed e.g. in 
\cite{lawler}.

While single-site starts are most efficient to locate the critical point \cite{torre},
and were used for this also in the present work, they cannot so easily be used for 
measuring persistence. For the latter, starts with a finite density of active sites
(say $\rho(0)=1/2$) are most straightforward, but they require enormous memory in
high dimensions, since hashing is of no use. If we want to have no finite-size 
effects, we have to use lattices whose linear size is larger than $t^{1/z}$, where 
$t$ is the simulation time and $z$ is the dynamic critical exponent which for DP
is $z=\nu_\|/\nu_\perp <2$. Even if we use multispin coding (i.e. 1 bit per site), we 
can only simulate rather short times on present-day workstations.

A way out of this dilemma is to use the time-reversed process discussed in 
\cite{Hinrichsen,Fuchs}. The persistence probability $P(t)$ is defined 
as the probability that none of the space-time points $(i,t')$ for fixed $i$ and $0\leq t'<t$ is activated by any 
of the active sites on the initial hypersurface $(i',0)$. This means that if site
$i$ is still persistent at time $t$, there cannot be any path of active sites (an
analog argument holds also for bond percolation) connecting any of the sites $(i,t')$ 
to any $(i',0)$. But such paths, if they would exist, could also be followed 
in the opposite direction. Thus $P(t)$ is also equal to the probability that the 
cluster of sites activated by ``sources" on the line interval $\{(i,t'); -t\leq t'< 0\}$ 
does not survive to time 0.

This is implemented in the following recursive way (see Fig.~1). Let us denote by 
${\cal C}_t$ the cluster activated by the sites (``seed") $\{(i,t'); -t\leq t'< 0\}$, and 
by ${\cal H}_0$ the hyperplane $\{(i',0); 0<=i'<N\}$. We start with $t=1$, in which case 
${\cal C}_t$ is just the site $(i,-1)$ plus all sites with $t' \geq 0$ connected to it.
If ${\cal C}_t\cap {\cal H}_0$ is not empty, i.e. if the seed activates at least one site in 
${\cal H}_0$, we discard ${\cal C}_t$ and start a new run. Otherwise, we add the site
$(i,-t-1)$ to the seed and activate all sites connected to it {\it which are not already 
in ${\cal C}_t$}. The new cluster, which will be at least as large as ${\cal C}_t$ but 
in general not much larger, is ${\cal C}_{t+1}$ (or rather, due to the particularities of the 
lattices used in this paper, ${\cal C}_{t+2}$; see the discussion at the end of this section). 
This is iterated until either the cluster 
intersects ${\cal H}_0$ or until $t=t_{\rm max}$ is reached.

\begin{figure}
  \begin{center}
\psfig{file=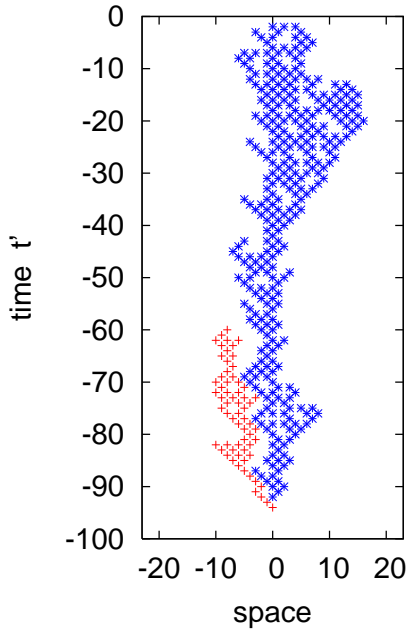,width=8.6cm, angle=270}
   \caption{(color online) Typical cluster of the time-reversed process, where a segment 
     of a line $i=0$, $-t \leq t' <0$ is the active seed. The growth is upward but in epochs,
     where each epoch corresponds to the downward extension of the seed by one site. The sites 
     activated during the last epoch are given a different color. Notice that only every
     second site (arranged in a checkerboard pattern) is used for the process. Notice also 
     that a straightforward growth of the cluster, without breaking it up into epochs, would
     be {\it much} less efficient. Typical clusters would be very fat, before applying the 
     conditioning of not cutting constant-$t$ hyperplanes, and most clusters would not
     meet the condition. The main virtue of the present algorithm is that such clusters 
     are eliminated already at a very early stage.}
   \label{fig1}
  \end{center}
\end{figure}

It might seem at first that this is much more storage demanding than direct simulation, 
since we have to store now the entire space-time bit pattern, not only the spatial 
pattern at a fixed time. What saves us, however, is that the clusters are fractal with 
spatial fractal dimension $\leq 2$, whence hashing will be very efficient. 

While $P(t)$ is the chance that ${\cal C}_t$ survives up to times $\geq t$, and thus 
$\theta$ describes its survival probability, the new critical exponent $\zeta$ describes 
the growth of its mass. There are several ways to describe this growth. The 
first, and maybe most natural would be the ansatz
\be
   M(t) \equiv \langle \# {\cal C}_t \rangle \sim t^{\zeta_1},
\ee
where $M(t)$ is the average mass of the still surviving clusters. Its main disadvantage 
is that contributions to $M(t)$ from early times might not scale, but decrease only 
slowly in importance as $t$ grows.

Alternatively we can define critical exponents via the increase of integrated 
quantities like $M(t)$, as this depends less on these non-scaling contributions.
Thus in the following we will define $\zeta$ via the average number 
$m(t)$ of activated sites (except for the site at the tip of the cluster) during the 
step ${\cal C}_{t-2}\to {\cal C}_t)$,
\be
   m(t) \sim t^\zeta.
\ee
If there were no correlation between the cluster mass and its survival probability, we 
would have obviously
\be
   \zeta_1 = \zeta+1.       \label{zeta1}
\ee
The same should still hold if there is exact scaling. Our numerics suggests that 
Eq.~(\ref{zeta1}) is correct, but with large finite-$t$ corrections. Appearant deviations
from Eq.(\ref{zeta1}) will dominate our error estimates.

In addition to its mass we can also measure other characteristics of the cluster, such 
as its spatial extension. As usual in time dependent critical phenomena we can define
$z'$ as 
\be
   R(t) \sim t^{1/z'}.
\ee
The same definition can also be used for ordinary DP clusters, where $z = \nu_\|/\nu_\perp$
\cite{Haye} \footnote{Unfortunately, in DP also $2/z$ is sometimes called $z$.}.
We measured $z'$ in $d=1$ and $d=2$ and found $z'=z$ within rather small error bars
($z' = 1.5803(9)$ for $d=1$, $z' = 1.770(9)$ for $d=2$; the corresponding values of $z$
are 1.5807 \cite{Jensen} and 1.767 \cite{Perlsman}). Thus we conjecture that $z'=z$  
exactly. The fact that critical exponents describing geometric aspects are more robust 
than others is also well known from standard critical phenomena,
where entropic boundary exponents are different from bulk exponents, 
while correlation length exponents do not change near boundaries \cite{Diehl}.

The relative efficiencies of direct simulations of half active lattices on the one hand, 
and the simulation of single time reversed clusters on the other hand, can be obtained
by estimating the number of sites that have to be tested/activated to establish that 
one site persists up to time $t$. In the direct approach this is 
\be
   n_{\rm tested}^{\rm direct} \approx \sum_{t'=0}^t \rho(t) / P(t) \sim t^{\theta+1-\delta},
\ee
where $\delta$ is the exponent for the decay of the density of active sites. For 
the time reversed cluster growth it is
\be
   n_{\rm tested}^{\rm reverse} \sim t^{\zeta+1}.
\ee
The ratio is
\be
   n_{\rm tested}^{\rm reverse} / n_{\rm tested}^{\rm direct} \sim t^{\zeta+\delta-\theta}.
   \label{efficiency}
\ee
If the exponent in this formula is negative, the time reversed cluster growth has less 
time complexity than the direct simulation.

Before leaving this section, let us make three remarks:
\begin{itemize}
\item For the estimates of $p_c$ we used the variance reduction method described in 
\cite{dp4d,percol}. This gave particularly big improvements for $d>4$.
\item As noticed already in \cite{Hinrichsen,Fuchs}, the regions left and right of the site
at which persistence is measured are decoupled and can be treated independently. Thus 
the sites $i'>0$ and $i'<0$ in Fig.~\ref{fig1} can be simulated independently. Both sides
contribute the same to $M(t)$ and to $\log P(t)$, so decay is much slowed down when only 
one half of the cluster is simulated and accuracy is substantially improved.
\item For any $d$ we use lattices where each bond changes only one of the spatial coordinates 
and, of course, time (more precisely, an active site can activate nearest neighbors in a 
simple hypercubic lattice at exactly one unit later time). Thus the lattices separate naturally 
into two checkerboard type sublattices. If activation is at the start restricted to one of 
them, it stays on it forever. Thus we can, without loss of generality, start with 
active sites restricted to one of the sublattices
(for $d=1$ this is clearly seen in Fig.~\ref{fig1}). Again this divides the exponent 
$\theta$ by a factor 2 as compared to simulations where both sublattices are active, and 
makes simulations more easy. The values of $\theta$ quoted below refer to activation 
restricted to one sublattice.
\end{itemize}

\section{Results}

Our main results are summarized in Table~1. Except for $d=1$, where extremely precise 
estimates of $p_c$ are available from series expansions \cite{Jensen}, we first made 
standard spreading simulations \cite{torre} where we measured the mass, survival probability,
and r.m.s radius of clusters grown from single point seeds. For $d=4$ we took into account 
the logarithmic corrections calculated in \cite{Janssen} in the same way as in \cite{dp4d}. 
Otherwise, we estimated the value of $p_c$ by demanding that the total number of all 
active sites is a pure power law for large $t$, up to possible corrections to scaling. Virtual 
lattice sizes were in each case $2^{64}$, which allowed for $t$ values at least twice as long 
as those used for measuring persistence. The results are given in the second column of Table~1.

\begin{table}
\begin{center}
\begin{tabular}{|r|l|l|l|r|c|c|} \hline
 $d$ & $\qquad p_c$ &  $\;\;\;\;\theta$ & $\;\;\;\;\zeta$ &  $t_{\rm max}\;\;$ & $N_{cl}/10^9$ & $\;\;\eta+\delta$ \\ \hline
  1  & .70548522(4) &  1.5167(7)      &  0.472(1)       &  195,000       &  19.5  &    0.47315\\   
  2  & .3445736(3)  &  1.611(7)       &  0.689(6)       &   19,360       &  85.0  &    0.685(6)\\
  3  & .2081040(4)  &  1.57(2) ?      &  0.85(1)        &    6,250       &  11.7  &    0.85(2) \\   
  4  & .1461593(2)  &  1.37(2)        &  0.97(2)        &    4,230       &   1.4  &    1.0 \\
  5  & .1123373(2)  &  1.216(12)      &  0.998(7)       &    3,130       &   1.0  &    1.0 \\   
  6  & .0913087(2)  &  1.175(11)      &  1.002(6)       &    2,610       &   .42  &    1.0 \\
  7  & .07699336(7) &  1.115(7)       &  1.002(4)       &    1,370       &   .40  &    1.0 \\ \hline
\end{tabular}
\caption{Main results. All values of $p_c$ are new estimates, except the value for $d=1$
which is from \cite{Jensen}. $t_{\rm max}$ is the maximum time over which clusters of 
the time-reversed process are grown, and $N_{cl}$ is the number of clusters
grown in this way, including those which died before reaching $t_{\rm max}$. For $d=2$
also a large number of direct simulations were made, typically up to the same 
$t_{\rm max}$, and their results are also taken into account in the estimate of  
$\theta$. The question mark for $\theta(d=3)$ indicates that the data are also compatible 
with some other relation (e.g. a stretched exponential) instead of a 
power law. The last column uses values from \cite{Haye}.}
\end{center}
\end{table}

\begin{figure}
  \begin{center}
\psfig{file=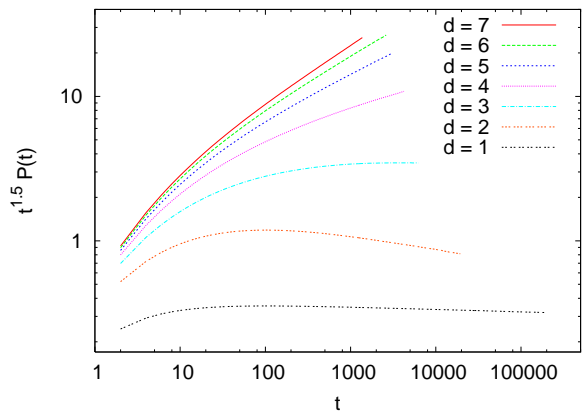,width=5.7cm, angle=270}
   \caption{(color online) Estimates of $t^{1.5}P(t)$ against $t$, for $d=1$ (bottom) to $d=7$ (top).
     Notice the very large corrections to scaling for $d=3$ which are indeed larger than
     those for $d=4$ where we expect logarithmic corrections.}
   \label{fig2}
  \end{center}
\end{figure}

Results for $P(t)$ are shown in Fig.~2. The curves for $d=2$ to 7 show the raw data (multiplied 
by $t^{1.5}$), while the curve for $d=1$ shows the square of the actually measured survival 
probability (see previous section). The data for $d=1$ show by far the smallest corrections 
to scaling -- they are well represented by a power law for $t>100$. In contrast, for $d=2$ and 
$d=3$ we see the extremely strong deviations from scaling noticed already in \cite{Fuchs}. 
A priori, the strongest deviations would have been expected for $d=4$ since this is the 
upper critical dimension for DP, and we must expect logarithmic corrections there as for all 
other observables. It is completely unclear why the corrections to scaling are largest for
$d=3$ instead, and why they also seem to decrease slowest with $t$ (they seem better 
compatible with logarithmic corrections than with power-law terms). Indeed, the data for 
$d=3$ taken by themselves might suggest a different relation between $\theta$ and $t$ than 
a power law, e.g. a stretched exponential. Although extremely unlikely from a theoretical 
point of view, we cannot really exclude this possibility. For $d>4$ the 
corrections to scaling are still large, but exponents defined for them via
\be
   P(t) \sim t^{-\theta} \left(1+{a\over t^\Delta}\right)   \label{delta}
\ee 
seem to increase. Thus, in spite of the visible curvature of all curves in Fig.~2, 
determination of $\theta$ becomes more reliable for larger $d$, as soon as $d\geq 5$.

Values of $\theta$, with subjective error bars dominated by the uncertainties of parameterizing 
the scaling corrections, are shown in Table~1. While our estimates of $\theta$ are typically 
larger than those of \cite{Albano}, they agree with those of \cite{Fuchs} with one minor 
exception: While $\theta > 1.62$ is quoted in \cite{Fuchs} for the contact process in 
$d=2$ (and extrapolating the corresponding curve in Fig.~3 of their paper would suggest
indeed a value {\it substantially} larger than 1.62), we obtain $1.611(1)$ for $d=2$. We 
believe that this is most likely due to an inaccurate value of the critical rate $\lambda_c$
used in \cite{Fuchs}.  

Table~1 suggests that $\theta\to 1$ in the limit $d\to\infty$, as predicted analytically 
in \cite{Fuchs}. But the agreement is not quantitative, as our values are much larger 
than those given by Eq.~(14) of \cite{Fuchs}. It might be that this is because the latter
was derived for bond percolation, but this seems unlikely in view of the good agreement 
between exponents obtained for (site) DP and for the contact process. This agreement 
strongly suggests universality.

\begin{figure}
  \begin{center}
\psfig{file=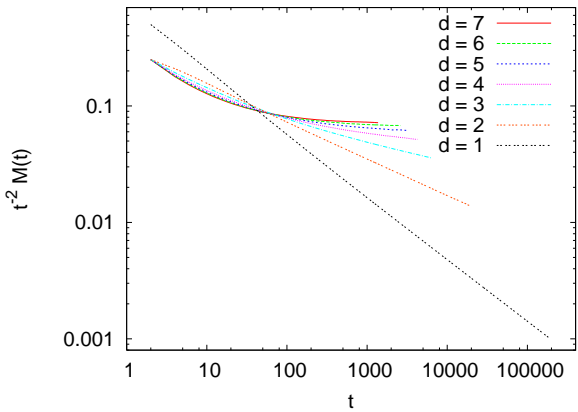,width=5.7cm, angle=270}
   \caption{(color online) Estimates of $M(t)/t^2$ against $t$, for $d=1$ (bottom) to $d=7$ (top).
     Here $M(t)$ is the number of active space-time sites (excluding the center line) in 
     inverse dynamics clusters ${\cal C}_t$.}
   \label{fig3}
  \end{center}
\end{figure}

\begin{figure}
  \begin{center}
\psfig{file=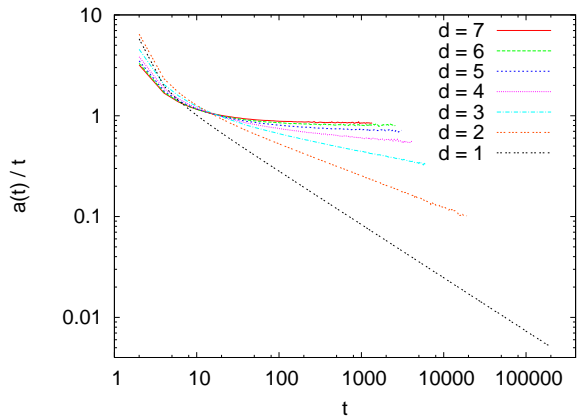,width=5.7cm, angle=270}
   \caption{(color online) Estimates of $m(t)/t$ against $t$, for $d=1$ (bottom) to $d=7$ (top).
     Here $m(t)$ is the number of active sites (excluding the center line which is always active)
     added to the inverse dynamics clusters ${\cal C}_t$ during the step $(t-2)\to t$. To reduce 
     statistical fluctuations, data have been binned with $\Delta t/t \approx 0.03$.}
   \label{fig4}
  \end{center}
\end{figure}

Log-log plots of the size $M(t)$ of inverse dynamics clusters ${\cal C}_t$ against $t$ are shown 
in Fig.~\ref{fig3}. To 
stress that $\zeta_1 = 2$ within statistical errors for $d>4$, and to reduce the plotting range 
on the $y$-axis, we plotted actually $M(t)/t^2$. Notice also that points on the central line
(which belong to ${\cal C}_t$ trivially) are excluded from $M(t)$. We see strong deviations from 
pure power laws for all dimensions $\neq 2$, but the data are in all cases compatible with 
asymptotic power laws. Similarly large deviations from pure power laws are also seen in log-log 
plots of $m(t)$ versus $t$ (see Fig.~\ref{fig4}), but the correction to scaling exponents
$\Delta$, defined via Eq.~(\ref{delta}), are systematically larger there. Thus extracting
asymptotic power laws from Fig.~\ref{fig4} is easier than from Fig.~\ref{fig3}, in spite of
larger statistical fluctuations. The asymptotic exponents extracted 
from Figs.~\ref{fig3} and \ref{fig4} should be related by Eq.(\ref{zeta1}). For all dimensions
this relation is roughly fulfilled, with the resulting estimates quoted in Table~1. The errors
quoted there are mostly reflecting the discrepancies between the estimates based on 
Figs.~\ref{fig3} and \ref{fig4}.

After submission of this paper, it was pointed out to me by Haye Hinrichsen \cite{HH} that 
$\zeta$  can be related to standard exponents. If we assume that the density in the time reversed
cluster decays radially, from a value near unity, as $r^{-z\delta}$ with a radial cutoff at 
$r_c \sim t^{1/z}$, then
\be 
   \zeta = d/z - \delta\;.                     \label{zet1}
\ee
Using the hyperscaling relation $d/z = 2\delta+\mu$ for $d\leq 4$, where $\eta$ describes the 
average number of sctive sites in single-site seeded DP clusters \cite{Haye}, this can be 
written as 
\be
   \zeta = \eta+\delta\;.                      \label{zet2}
\ee
While Eq.~(\ref{zet1}) can only hold for $d\leq 4$ (for $d>4$ the assumption of a power law 
radial density decay from a value of order unity must fail), Eq.~(\ref{zet2}) seems to hold
for all dimensions. To demonstrate this, we quote in the last column of Table~1 the values 
of $\eta+\delta$ from \cite{Haye}. In all cases, agreement holds to better than two standard
deviations. Notice that Eq.~(\ref{zet2}) is very surprising. It says
that two cluster growth mechanisms which seem to have no close relationship, and for which 
the cluster survival probabilities scale with different critical exponents, have nevertheless 
the same growth exponents after conditioning on survival.

Finally, let us discuss the relative time complexities of direct and time reversed (single
cluster) simulations using Eq.(\ref{efficiency}) and the critical exponents given in Table~1 and
in \cite{Haye}. For $d=1$ we find that the time reversed simulation is much more efficient, by
a factor $\sim t^{0.9}$. Thus one gains several orders or magnitude in CPU time. The same is
qualitatively true for $d=2$, although the difference is much less, only a factor $\sim t^{0.4}$.
For $d= 3$ both methods have the same time complexity, and for $d>3$ the ranking is reversed --
quite substantially ($t^{-0.9})$ for $d=7$. Thus direct simulations would be much faster in
high dimensions, if we could solve the space complexity (memory limitation) problem. The other
advantage of time reversed simulations is of course that they offer the unique possibility to
measure the exponent $\zeta$.

\section{Discussion}

We presented in this paper numerical estimates for the local persistency exponent $\theta$ in 
DP. These estimates are more precise than previous ones for spatial dimensions $d \leq 4$, and 
are the first published ones for $d>4$. The latter were made possible by a combination of 
hashing (virtual memory) and a novel way to simulate the growth of the time reversed clusters 
discussed in \cite{Hinrichsen,Fuchs}. The latter algorithm suggested also very naturally
another new critical exponent, $\zeta$. It seems that $\theta$ and $\zeta$ are not 
trivially related, but that $\zeta$, which describes the cluster growth of an artifical 
time-reversed evolution with modified dynamics, is the same as the usual DP exponent describing 
mass growth of active clusters. While this is supported both by an indirect heuristic 
argument and by numerics, it seems very surprising and not well understood intuitively. 

We verified the very large corrections to scaling seen in previous analyses local persistency
in DP, in particular for 2 and 3 spatial dimensions. Indeed, these corrections seem to be 
larger in $d=3$ than in $d=4$, where we would have expected a priori that logarithmic 
corrections should give the strongest deviations from a pure power law. Although we see 
no theoretically appealing scenario for it, the data alone would suggest that the asymptotic
behavior of $P(t)$ in three dimensions is not described by a power law at all, but rather 
by a stretched exponential or something similar.

In the present paper we dealt only with directed percolation, but very similar problems 
can be studied also for ordinary (undirected) percolation. The situation is maybe best 
described in two dimensions (analogous to 1 spatial plus 1 temporal dimension in DP). 
Let us consider a wedge cut out from a square lattice, with edges $W_1$ and $W_2$ meeting in 
point $P$ and extending to infinity away from $P$. The angle between $W_1$ and $W_2$ is 
$\alpha$. Let us furthermore consider intervals
$I_1$ and $I_2$ on $W_1$ resp. $W_2$, with lengths $\ell_1$ and $\ell_2$. We 
can then ask for the probability $P(I_1,I_2,\alpha)$ that there is no path from any point 
on $I_1$ to any point on $I_2$. This is complementary to the probability that there is a 
path from $I_1$ to $I_2$. Probabilities of the latter type (called `crossing probabilities') 
were studied extensively \cite{Langlands,Cardy,Kleban}, but it seems that 
neither $P(I_1,I_2,\alpha)$ nor its generalizations
to higher dimensions were studied previously. 

Acknowledgments: I am indebted to Purusattam Ray for introducing me to this subject 
and for pointing out Ref.~\cite{Fuchs}. Without his encouragement and insistence I would
not have started this work. I also want to thank the Chennai Institute of Mathematical Sciences,
where this work was begun, for its hospitality and stimulating atmosphere. But most of all 
I am indebted to Haye Hinrichsen for pointing out to me Eq.~(\ref{zet1}), and for allowing 
me to use this unpublished result in the present paper.


\begin{thebibliography}{99}
\bibitem{Bray} A.J. Bray, B. Derrida, and C. Godr\`eche, Europhys. Lett. {\bf 27}, 175 (1994).
\bibitem{Majumdar} S.N. Majumdar, Curr. Sci. India {\bf 77}, 3704 (1999).
\bibitem{Ray} P. Ray, Phase Trans. {\bf 77}, 563 (2004).
\bibitem{Fuchs} J. Fuchs, J. Schelter, F. Ginelli, and H. Hinrichsen,  J. Stat. Mech. P04015 (2008).
\bibitem{Hinrichsen} H. Hinrichsen and H.M. Koduveli, Eur. Phys. J. B {\bf 5}, 257 (1998).
\bibitem{Albano} E.V. Albano and M.A. Mu\~noz, Phys. Rev. E {\bf 63}, 031104 (2001).
\bibitem{HH} H. Hinrichsen, private communication.
\bibitem{Haye} H. Hinrichsen, Adv. Phys. {\bf 49}, 815 (2000).
\bibitem{lawler} P. Grassberger, arXiv:0905.3440 (2009).
\bibitem{torre} P. Grassberger and A. de la Torre, Ann. Phys. {\bf 122}, 373 (1979).
\bibitem{Jensen} I. Jensen, J. Phys. A: Math. Gen. {\bf 32}, 5233 (1999).
\bibitem{Perlsman} E. Perlsman and S. Havlin, Europhys. Lett. {\bf 58}, 176 (2002).
\bibitem{Diehl} H.-W. Diehl, {\it Phase Transitions and Critical Phenomena} Vol. 10, 
   C. Domb and J.L. Lebowitz (Eds.) (Academic Press, New York 1986).
\bibitem{dp4d} P. Grassberger, arXiv:0904.0804 (2009).
\bibitem{percol} P. Grassberger, Phys. Rev. E {\bf 67}, 036101 (2003).
\bibitem{Janssen} H.-K. Janssen and O. Stenull, Phys. Rev. E {\bf 69}, 016125 (2004).
\bibitem{Langlands} R. Langlands, C. Pichet, P. Pouliot, and Y. Saint-Aubin, J. Stat. Phys. {\bf 67} 553 (1992).
\bibitem{Cardy} J. Cardy, J. Phys. A: Math. Gen. {\bf 25}, L201 (1992).
\bibitem{Kleban} P. Kleban, Physica A {\bf 281}, 242 (2000).

\end{thebibliography}
\end{document}